# Magnetocaloric effect in $Ni_2(Mn,Cu)Ga_{0.84}Al_{0.16}$ Heusler alloys

A.A. Mendonça[1], L. Ghivelder[1], J.F. Jurado[1,2], and A.M. Gomes[1]

[1] Instituto de Física, Universidade Federal do Rio de Janeiro, C.P. 68528, Rio de Janeiro 21941-972, Brazil

[2] Lab. Propiedades Térmicas Dieléctricas de Compositos, Depto. de Física y Química, Universidad Nacional de Colombia, A.A 127 Manizales, Colombia

## Abstract

Polycrystalline Heusler compounds $Ni_2Mn_{0.75}Cu_{0.25}Ga_{0.84}Al_{0.16}$ with a martensitic transition between ferromagnetic phases and $Ni_2Mn_{0.70}Cu_{0.30}Ga_{0.84}Al_{0.16}$ with a magnetostructural transformation were investigated by magnetization and thermal measurements, both as a function of temperature and magnetic field. The compound $Ni_2Mn_{0.75}Cu_{0.25}Ga_{0.84}Al_{0.16}$ presents a large magnetocaloric effect among magnetically aligned structures and its causes are explored. In addition, $Ni_2Mn_{0.70}Cu_{0.30}Ga_{0.84}Al_{0.16}$ shows very high, although irreversible, entropy and adiabatic temperature change at room temperature under a magnetic field change 0-1 T. Improved refrigerant capacity is also a highlight of the 30% Cu material when compared to similar $Ni_2MnGa$-based alloys.

## Introduction

A magnetostructural transformation, the coupling of structural and magnetic ordering transitions, often increases the magnetocaloric and ferromagnetic shape memory potential of some compounds due to the considerable magnetic field dependence of the magnetostructural transition [1]. Due to this coupling, a high total entropy change ($\Delta S$) can be achieved with a structural transformation concomitant with magnetic transition between a magnetically ordered phase (ferromagnetic) and a magnetically misaligned phase (paramagnetic or antiferromagnetic). Among the main magnetocaloric materials which present magnetostructural transitions, Heusler alloys based on Ni-Mn-X (X = Ga, In, Sn and Sb) have received particular attention [2]. Some ferromagnetic $Ni_2MnGa$-based Heusler alloys are well known due to interesting properties, such as martensitic transformations [3] and large magnetic-field-induced deformations due to direct or reverse martensitic transitions [4,5,6] as well as reorientation of the martensite variants by twin boundary motion [7,8,9]. These materials also present an enhanced magnetocaloric effect (MCE) in some off-stoichiometry compounds [10,11,12].

$Ni_2MnGa$ crystallizes in a cubic $L2_1$-type structure (space group Fm-3m) with a room temperature lattice parameter a = 5.825 Å and a low temperature martensitic structure [3]. It shows a continuous transition between a ferromagnetic martensite phase and a paramagnetic austenite phase at 376 K, as well as a martensitic transition around 200 K [3]. Small additions of Ni in the Mn site lead to the appearance of a magnetostructural transition when both structural and magnetic transitions are very close in temperature, in this case around 333 K [10]. The possibility to have these transitions around the same temperature was also reported in $Ni_2Mn_{1-x}Cu_xGa$ alloys, where values of entropy change as large as -60 $J/Kg^{-1}K^{-1}$ were observed for x ≈ 0.25 with a 0 - 5 T field change [11,13,14,15].

The fabrication method and annealing processes are important factors contributing to the alloy's magnetic and structural properties, as well as ensuring predominance of the ferromagnetic $L2_1$-type structure [16]. Aluminium addition in $Ni_2MnGa_{1-x}Al_x$ alloys yields a coexistence of $L2_1$ (ferromagnetic) and B2 (antiferromagnetic) structures, leading to a predominant antiferromagnetism when x > 0.30 [16,17]. In $Ni_2Mn_{1-x}Cu_xGa_{0.9}Al_{0.1}$, with 10% of Al substitution on the Ga site [18], the crystallographic and magnetic transitions coexist at around 295 K for x = 0.20, resulting in ΔS = - 9.5 $JKg^{-1}K^{-1}$ under a 0 - 5 T magnetic field change. Although Al substitution drastically reduces the maximum of ΔS when compared to the Al free alloy, the refrigerant capacity, RC ~ 110 J/Kg for x = 0.20, is larger than for $Ni_2Mn_{0.75}Cu_{0.25}Ga$ [11].

A very recent investigation correlates thermal hysteresis and phase compatibility as the Cu content is varied in the Heusler alloys $Ni_2Mn_{1-x}Cu_xGa_{0.84}Al_{0.16}$ [19]. In addition, a previous study on $Ni_2Mn_{0.70}Cu_{0.30}Ga_{0.84}Al_{0.16}$ focused on structural and magnetostriction measurements, which revealed a large magnetic field induced strain under low fields [20] which might be relevant for applications related to shape memory effects. In the present work, we studied the magnetocaloric properties of the materials $Ni_2Mn_{0.75}Cu_{0.25}Ga_{0.84}Al_{0.16}$ and $Ni_2Mn_{0.7}Cu_{0.3}Ga_{0.84}Al_{0.16}$. The latter is similar to the well-known $Ni_2Mn_{0.75}Cu_{0.25}Ga$, with partial Ga replacement by Al. This study is based on magnetization and heat flow measurements. Our results for entropy change, adiabatic temperature change, and refrigerant capacity are compared to some well-known MCE materials found in the literature.

## Experimental Methods

Two samples pellets of 1.5 g with composition $Ni_2Mn_{0.75}Cu_{0.25}Ga_{0.84}Al_{0.16}$ and $Ni_2Mn_{0.70}Cu_{0.30}Ga_{0.84}Al_{0.16}$ were made using conventional arc melting process in 99.999% pure argon atmosphere and metallic elements with purity better than 4N. The samples were re-melted 3 times, with care not to keep the arc for more than five seconds, thereby avoiding large losses due to Mn vaporization. Initially, Mn losses of approximately 3% at the end of the melting process were observed, and to account for this, we added an excess of Mn before melting to ensure the correct stoichiometry. To achieve a better sample homogenization, two thermal treatments were applied. The samples were wrapped with tantalum foil and encapsulated in quartz tubes under a low argon pressure of 0.2 atm. The first thermal treatment was done for 72 h at 1273 K and the second for 24 h at 673 K, both at a rate of 3 K/min and quenching in room temperature water at the end of each process. Isothermal and isofield magnetization measurements were made using a Vibrating Sample Magnetometer in a Physical Properties Measurement System (PPMS) from Quantum Design Inc. A Peltier differential scanning calorimeter device was built to measure heat flow under magnetic field in the PPMS platform. A similar apparatus is described in Ref. [21]. Additional heat flow measurements were made using a commercial Differential Scanning Calorimeter (DSC), model Q2000 from TA Instruments Inc, in order to compare the results with the home-made Peltier based DSC. The enthalpy change in the transition, obtained by the integrating the data of heat flow as a function of time, coincides within 2.2% when comparing the results of the Peltier Cell and the commercial DSC. The latter was also used for specific heat measurements.

**Results and Discussion**

The temperature dependence of the magnetization measured in ZFC (zero field cooled) and FCC (field cooled cooling) modes under a magnetic field of 20 mT is shown in Fig. 1. The compound with x = 0.25 presents a magnetic transition in the austenite phase with $T_C$ = 296 K on cooling. As the temperature decreases, a ferromagnetic martensite phase takes place from the ferromagnetic austenite one at 262 K, with thermal hysteresis around 6 K. Partial Ga replacement by Al at 16% separates the magnetostructural transformation of the $Ni_2Mn_{0.75}Cu_{0.25}Ga$ compound [11] into a first order martensitic and a second order magnetic transformation, with 33 K of temperature difference.

For x = 0.30, the martensite phase is ferromagnetic while the austenite phase is paramagnetic, due to a magnetostructural transition that starts around 297 K on cooling, with thermal hysteresis around 9 K. The different magnetic ordering between the structural phases is a relevant property for magnetocaloric and ferromagnetic shape memory materials because it tends to increase the magnetization difference between the phases in the transformation.

In addition, ZFC and FCC magnetization measured at higher magnetic fields, from 20 mT up to 5 T, are shown in Fig. 2. Here, we notice in panel (a) that the x = 0.25 sample presents martensitic transformation among two ferromagnetic phases. Due to larger magnetocrystalline anisotropy, associated to the variants of martensite that impose constrains to the structure [5,10], the martensite phase for x = 0.25 presents lower magnetization at small magnetic fields while the opposite is observed at high fields. Since the magnetization process of these materials is strongly related to the structure, this higher magnetocrystalline anisotropy imposes a barrier to the magnetization at low magnetic field, and a higher field must be applied to overcome this barrier. As seen in the Fig. 2 (a), a magnetic field of 0.3 T is enough for the martensite magnetization to surpass the austenite magnetization.

On the other hand, the behavior associated to the magnetocrystalline anisotropy is not observed for x = 0.30 because this material transforms from a paramagnetic austenite to a ferromagnetic martensite. In this case, the benefit of a para-ferromagnetic transition is the increase of the magnetization difference between the phases, which is much higher for x = 0.30 when compared to the x = 0.25 sample, as observed in Fig. 2. The higher the magnetization difference between the phases implies that a lower magnetic field is required to induce the transformation. This yields an increase of the energy that the material can release under an external stimulus due to the magnetic contribution, which favors the composition x = 0.30 in terms of magnetocaloric properties.

For both samples, isothermal magnetization measurements up to 9 T are shown in Fig. 3. Prior to each measurement the material was heated to the high temperature austenite phase and then cooled to the target temperatures displayed in Fig. 3. For $x = 0.25$, shown in panel (a), at 262 K and 270 K the material is in a ferromagnetic state for both martensite and austenite phases. All isotherms measured between these two temperatures display a magnetic-field-induced martensitic transformation with 7 $Am^2kg^{-1}$ magnetic saturation difference. Even surrounded by ferromagnetic phases, this saturation difference provides enough magnetic driving force to induce a structural change, leading to a sudden increase in the magnetization of the sample once a critical field ($H_C$) is achieved. The temperature dependence of $H_C$ (not shown) presents a linear behavior, with slope of 1.5 T/K. In inset of Fig. 3 (a), the energy loss as a function of temperature is shown. The calculated energy loss was obtained by the area between the curves of increasing and decreasing fields (the latter is not shown) for temperatures in which $H_C < 5$ T. Technically, this lost energy is what reduces the energy exchanged by the material in cyclical applications. Therefore, the knowledge of this parameter is useful to identify temperature ranges in which the magnetic material is more efficient for application purposes.

In the results for the sample $x = 0.30$, seen in Fig. 3 (b), the data at 293 K shows a ferromagnetic-like curve, while at 306 K it exhibits a predominant paramagnetic behavior. The measurements from 298 K to 304 K reveal that the magnetostructural transition is induced by the magnetic field. The magnetic saturation difference between martensite and austenite is $\approx$ 17 $Am^2kg^{-1}$. Similar to the $x = 0.25$ sample, the critical field for the $x = 0.30$ compound also varies linearly with temperature (not shown), and the slope of the critical field is 1.2 T/K. As noticed in Fig. 2, the magnetization difference between the phases is larger for $x = 0.30$ than for x = 0.25. The energy loss increases linearly from 297 K to 302 K, as shown in the inset of the Fig. 3 (b) for 0 - 5 T magnetic field change. It is almost zero at 297 K and reaches its maximum at 302 K with ≈ 88 J/kg.

Both structural phases are ferromagnetic for the x = 0.25 compound. Nevertheless, due to the higher magnetic saturation of the martensite phase, the transition temperature increases under an applied field, as seen in Fig. 2. Quantitatively, the transition temperature shifts due to the action of the magnetic field with 0.65 K/T. On the other hand, the composition x = 0.30 with a para-ferromagnetic structural phase change has a higher magnetic driving force due to the larger magnetization difference between the phases. The transition temperature shifts with an applied field with a rate of 1.0 K/T. Theoretically, a thermodynamical model from Clausius-Clapeyron for first-order transformations [22,23] predicts that the magnetic-field-induced transition temperature shift is given by $dT/d(\mu_0 H) = -\Delta M/\Delta S$, where $\mu_0$ is the permeability of the free space, H is the magnetic field, ΔM and ΔS are the difference of magnetic saturation and the entropy change between the austenite and martensite phases. For this analysis, the values of ΔS were calculated from the ratio $Q/T_M$, where the latent heat Q under zero field was obtained from Ref. 19, and the transition temperature $T_M$, also under zero field, extracted from our magnetic measurements. Following this expression, the predicted transition temperature shift is 0.5 K/T for x = 0.25 and 0.8 K/T for x = 0.30, both only ~20% smaller than the measured values.

Both compositions display a first order transformation, with energy loss due to irreversibility. The amount of energy dissipated is related to the hysteresis of the phase change. When the thermal hysteresis is considerably larger than the transition temperature width $T_S - T_F$ , where $T_S$ and $T_F$ are the start and final temperatures of the first order transformation, a large magnetic driving force is needed to overcome the hysteretical barrier. The martensitic transition for the composition x = 0.25 has a temperature width $T_S - T_F \approx 2.3$ K, and a thermal hysteresis ≈ 6 K. For x = 0.30, the magnetostructural transformation has a temperature width ≈ 3.1 K and thermal hysteresis ≈ 9 K. Then, the ratio between thermal hysteresis and transition temperature width is 2.6 for x = 0.25 and 2.9 for x = 0.30. From this analysis, we understand that the transition is abrupt yet considerably hysteretical. This type of behavior favors a large strain [20] and large magnetocaloric effect under low magnetic fields, but reversibility is not expected since the transition temperature change under the action of magnetic field is considerable smaller than the thermal hysteresis.

In order to quantify the total entropy change ΔS of the samples under a magnetic field change, heat flow measurements were performed using a Peltier Cell as a thermal probe. This experimental set up allow us to measure the thermoelectric voltage associated with the energy released or absorbed by the material in an isothermal process, while the transformation is induced by the magnetic field.

Examples of the measured voltage data are show in Fig. 4, for both x = 0.25 and x = 0.30 samples. These curves were obtained with different protocols of magnetic field change, while the probe measures the heat exchange. In the first one, show in Fig. 4 (a) for x = 0.25, the magnetic field varies in two steps, for 0 to 2 T and from 2 to 5 T. In the next case, displayed in Fig. 4 (b) corresponding to the x = 0.30 sample, the magnetic field changes continuously from 0 to 5 T. Finally, in Fig. 4 (c), again for x = 0.30, the magnetic field is changed in several discrete steps. The satellite peaks of heat released by the samples as the magnetic field varies in all parts of Fig. 4 signals that the transformation for both samples takes place in multiple steps.

From each isothermal measurement, we extract the heat exchanged and calculate ΔS [**Erro! Indicador não definido.**], as shown in Figs. 5 (a) and (b) as a function of temperature and magnetic field. The maximum ΔS and RC values obtained for 0 - 2 T and 0 - 5 T are summarized in Table 1. The RC values were calculated by integrating the ΔS(T) peak at half of the maximum height. Also in Figs. 5 (a) and (b), we plotted the entropy change, $\Delta S_M$, for 0 - 2 T calculated from the magnetization data [24]. In the case with a magnetostructural transformation, for x = 0.30, $\Delta S_M$ has a maximum value very close to the maximum of ΔS obtained from calorimetric data. On the other hand, for x = 0.25 the maximum value of $\Delta S_M$ is considerably lower than the maximum of ΔS from calorimetric data.

The difference in ΔS when comparing values obtained from magnetization and calorimetric results is related to the magnetic nature of the transformation. In the x = 0.25 sample, with martensitic transition, both phases are ferromagnetic, therefore a considerable part of the transition enthalpy comes from the lattice rearrangement, which has a non-magnetic origin. Since both phases are magnetically ordered the magnetic entropy change has a small contribution to the total ΔS. On the other hand, for x = 0.30, with a magnetostructural transformation, the magnetic contribution to the total entropy change is large, since the parent phase is paramagnetic and the resulting phase is ferromagnetic. Therefore, ΔS values obtained from magnetization data approaches the total entropy change obtained from thermal measurements.

The value of the zero magnetic field enthalpy for the martensitic transition of the x = 0.25 sample is ≈ 3630 J/kg [19]. This enthalpy yields a total entropy change under zero field of ≈ 14 $Jkg^{-1}K^{-1}$. However, the maximum entropy change under 0 - 5 T field change is ≈ 21 $Jkg^{-1}K^{-1}$. Then, even though both phases are ferromagnetic and we expect a small magnetic entropy change, there is still some considerable magnetic contribution within the ferromagnetic phases since the ΔS increased 50%.

The Bean and Rodbell model [25] of magneto-elastic coupling is useful to treat the magnetocaloric effect in a first order transformation by considering different contributions to the total entropy change. The model describes the total entropy in a first order transition as composed by three parts:

$$s = s_M(m) + s_W(m) + s_S(p,T)$$

where $s_M(m)$ is the magnetic entropy, $s_W(m)$ is the structural lattice entropy and $s_S(p,T)$ is the magneto-elastic entropy, related to the structure, that appears from ferromagnetic exchange forces in a magneto-elastic interaction. The details associated to each term are given in Ref. [26]. When the total entropy varies, we may have a magnetic contribution to the total entropy change that is not a magnetic entropy change; it is instead a magneto-elastic entropy change. This gives an insight about why the total entropy change in the x = 0.25 compound presents a considerable magnetic contribution although the martensitic transformation occurs between ferromagnetic phases.

It is relevant to compare ΔS and RC in our x = 0.30 sample with values obtained for similar materials in the literature. In $Ni_2Mn_{0.75}Cu_{0.25}Ga$ [11] and $Ni_{50}Mn_{18.5}Cu_{6.5}Ga_{25}$ [13], extremely large values of entropy change were reported, twice as high as the ones obtained here for 0 - 5 T change. Indeed, Ga replacement by Al decreases the maximum ΔS [18]. On the other hand, RC = 84 J/kg for $Ni_2Mn_{0.75}Cu_{0.25}Ga$ [11] and 94.6 J/kg for $Ni_{50}Mn_{18.5}Cu_{6.5}Ga_{25}$ [13], while our material with x = 0.30 presents RC = 120 J/Kg. Therefore, adding Al in the alloys decreases the maximum of ΔS, but increases the full width at half maximum (FWHM) of the ΔS(T) peak, leading to a higher RC. When compared to $Ni_2Mn_{0.8}Cu_{0.2}Ga_{0.9}Al_{0.1}$ of Ref. [18], we observe that our material, with more Al, presents considerably larger ΔS and RC. Our values of ΔS and FWHM are intermediate among those reported for $Ni_2Mn_{1-x}Cu_xGa$ and $Ni_2Mn_{1-x}Cu_xGa_{0.9}Al_{0.1}$ alloys, but the combination of both parameters, in our case, results in a maximized RC.

The magnetic field dependence of ΔS for x = 0.25 at 264 K and x = 0.30 at 298 K are shown in the insets of Figs. 5 (a) and (b), respectively. At these temperatures, around 90% of the ΔS for 0 - 5 T is achieved with only 0 - 1.5 T field change. This low field saturation of the entropy change at those temperatures is achieved due to the abrupt nature of the transformation. For example, for x = 0.30 and at $\mu_0H$ = 1 T, the material presents ΔS = -15 $Jkg^{-1}K^{-1}$. The materials under study in this work present ΔS values larger than the compound with Al content 0.10 [18] due to their sharper transformation. These high values of ΔS under low magnetic field change (0 - 1 T), are compatible with the large strain under low magnetic fields previously observed [20]. For comparison, the values of ΔS for our materials as well as for other compounds in the literature with large ΔS under 0 - 1 T are compiled in Table 2.

In order to calculate the adiabatic temperature change ($\Delta T_{AD}$), we measured the specific heat for both studied samples (not shown). These results combined with magnetization data of Fig. 3 allow us to calculated $\Delta T_{AD}$ through the expression [27]

$$\Delta T_{AD}\,(T,\Delta H) = -\int_{H1}^{H2} \frac{T}{C_{H,P}} \left(\frac{\partial M}{\partial T}\right)_{H,P} dH$$

where T is the temperature, H is the magnetic field, $C_{H,P}$ is the specific heat at constant magnetic field and pressure, and M is the magnetization.

The $\Delta T_{AD}$ curves for x = 0.25 and 0.30 under field changes of 0 - 1 T, 0 - 2 T and 0 - 5 T are exhibited in the Fig. 6. To avoid an overestimation due to first order transition and spikes, we used the methods described in Refs. [24] and [28]. Under a magnetic field change of 0 - 1 T we obtained the maximum of $\Delta T_{AD} \approx$ 1.0 K and 4.8 K for the compounds with x = 0.25 and 0.30 respectively. A comparison with values found in literature displayed in Table 2 shows that the composition x = 0.30 presents a high value of $\Delta T_{AD}$ for such magnetic field change. However, it should be mentioned that direct measurements of $\Delta T_{AD}$ would be preferable for this analysis.

## Conclusions

Magnetocaloric properties of the $Ni_2Mn_{1-x}Cu_xGa_{0.84}Al_{0.16}$ Heusler alloys for x = 0.25 and 0.30 were studied. The materials present a ferromagnetic martensite phase that evolves from ferromagnetic austenite for x = 0.25 and from paramagnetic austenite for x = 0.30. In addition to the ferromagnetic shape memory effect previously reported for x = 0.30, we show that these samples also display a large magnetocaloric effect, with an entropy change ΔS = -14 $JKg^{-1}K^{-1}$ at the martensitic transformation for x = 0.25, and ΔS = -21 $JKg^{-1}K^{-1}$ at the magnetostructural transition for x = 0.30, both under 0 - 2 T magnetic field change. When compared to the previous studied compounds $Ni_2Mn_{0.75}Cu_{0.25}Ga$ and $Ni_2Mn_{0.8}Cu_{0.2}Ga_{0.9}Al_{0.1}$, our material x = 0.30 presents a higher refrigerant capacity, achieving RC = 120 J/Kg for 0 - 5 T field change. A comparison with well know magnetocaloric materials with large values of ΔS and $\Delta T_{AD}$ around room temperature for 0 - 1 T magnetic field change shows that $Ni_2Mn_{0.7}Cu_{0.3}Ga_{0.84}Al_{0.16}$ presents impressive magnetocaloric properties under relatively low magnetic fields. However, although the sharp transition and magnetization difference between the phases yield good magnetocaloric properties, the results are not reversible and the material needs to be heated to the parent phase to undergo the transformation again.

**Acknowledgements**

This work was partially supported by the Brazilian agencies FAPERJ (Projects E-26/202.820/2018 and E-26/010.101136/2018) and CNPq (Project 424688/2018-2). A.A.M. was supported by a graduate fellowship from CAPES. We acknowledge helpful discussions with Prof. Lesley Cohen from Imperial College, UK.

**Figures and Tables:**

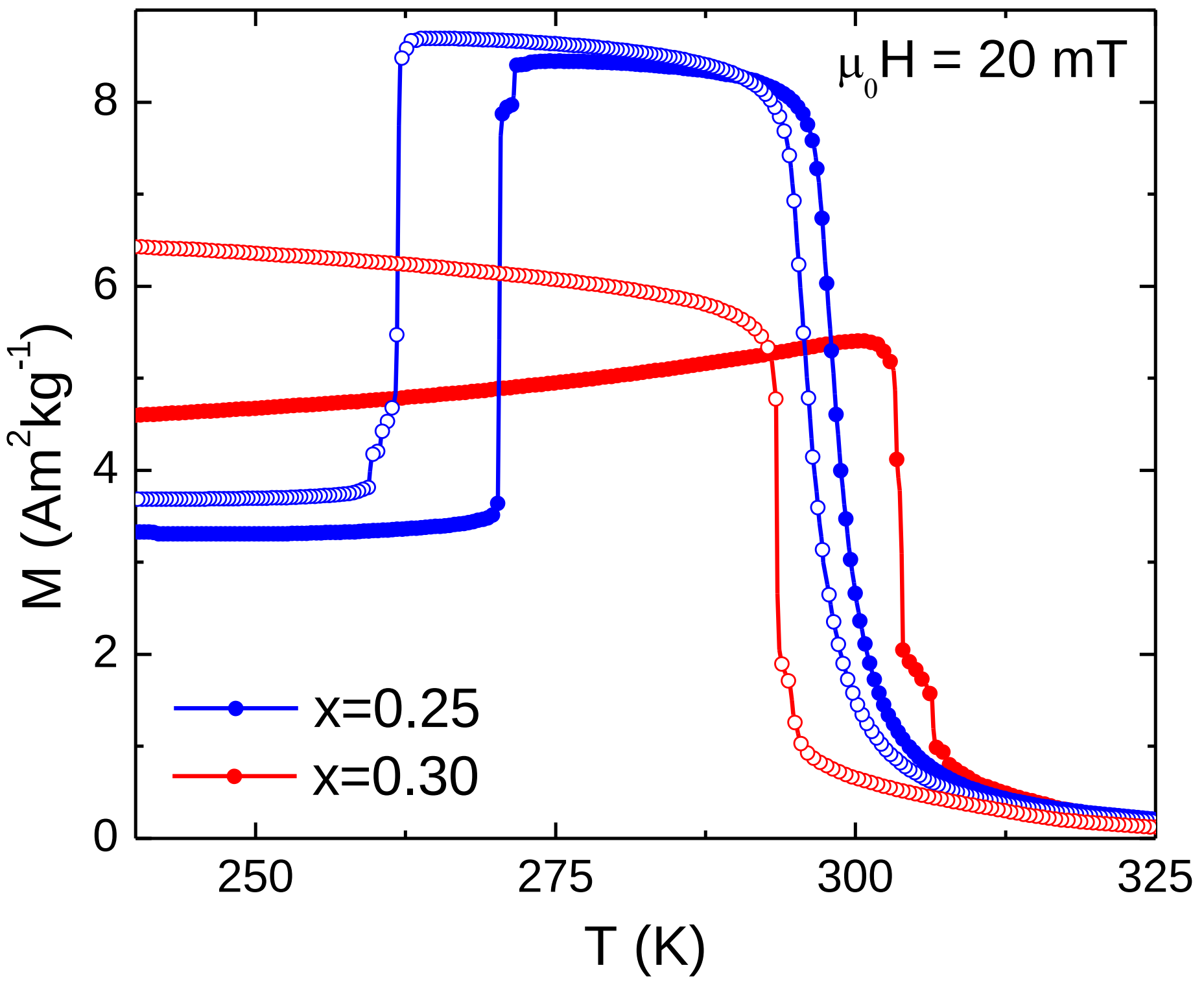

Figure 1: Temperature dependence of ZFC (full symbols) and FCC (open symbols) magnetization of $Ni_2Mn_{1-x}Cu_xGa_{0.84}Al_{0.16}$ under 20 mT field. The sample with x = 0.25 presents magnetic and martensitic transitions while the compound with x = 0.30 presents magnetostructural transformation at room temperature.

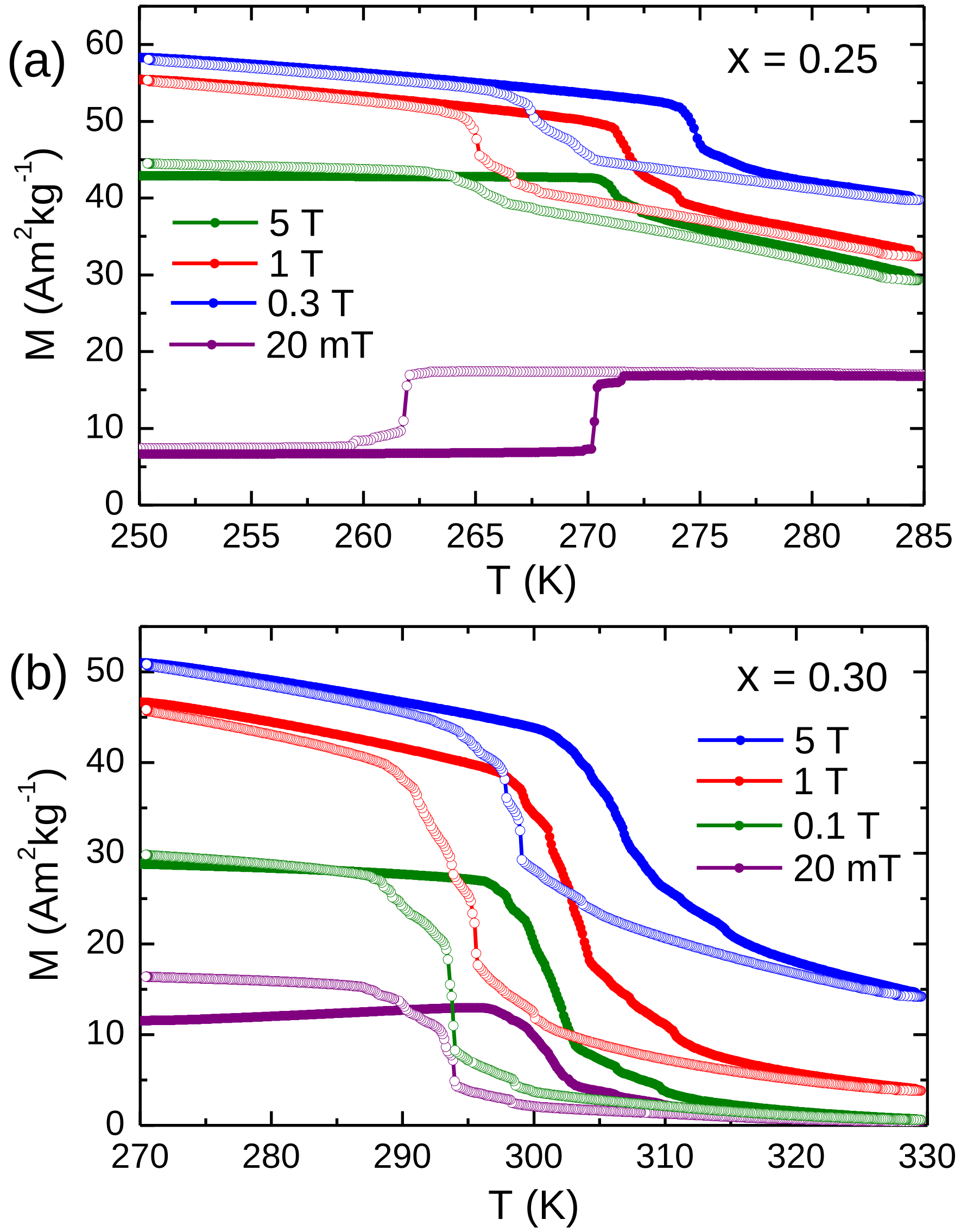

Figure 2: Magnetization as a function of temperature of $Ni_2Mn_{1-x}Cu_xGa_{0.84}Al_{0.16}$ for several magnetic fields. For x = 0.25 (a), the ferromagnetic martensite phase has a lower magnetization than the ferromagnetic austenite phase at 20 mT, and higher magnetization for fields above 0.3 T. On the other hand, the material x = 0.30 (b) presents a transformation from ferromagnetic martensite to paramagnetic austenite. The data for 20 mT was multiplied by two to improve visualization.

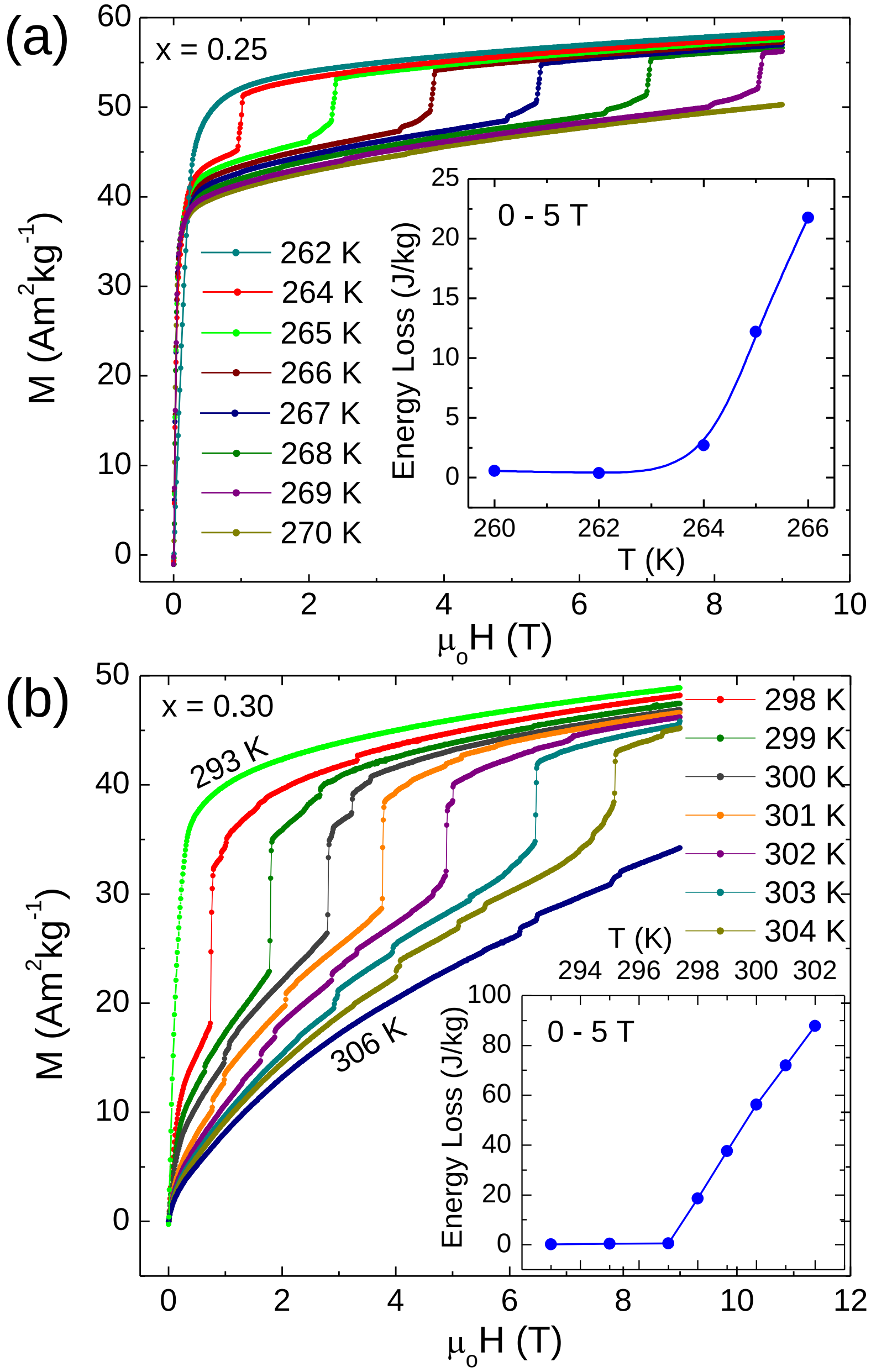

Figure 3: Magnetization as a function of magnetic field of $Ni_2Mn_{1-x}Cu_xGa_{0.84}Al_{0.16}$ ;(a) x = 0.25 and (b) x = 0.30. Both martensitic (x = 0.25) and magnetostructural (x = 0.30) transitions are induced by magnetic field changes. Inset: temperature dependence of the energy loss for 0 – 5 T calculated by increasing and decreasing (not shown) magnetization for each isotherm.

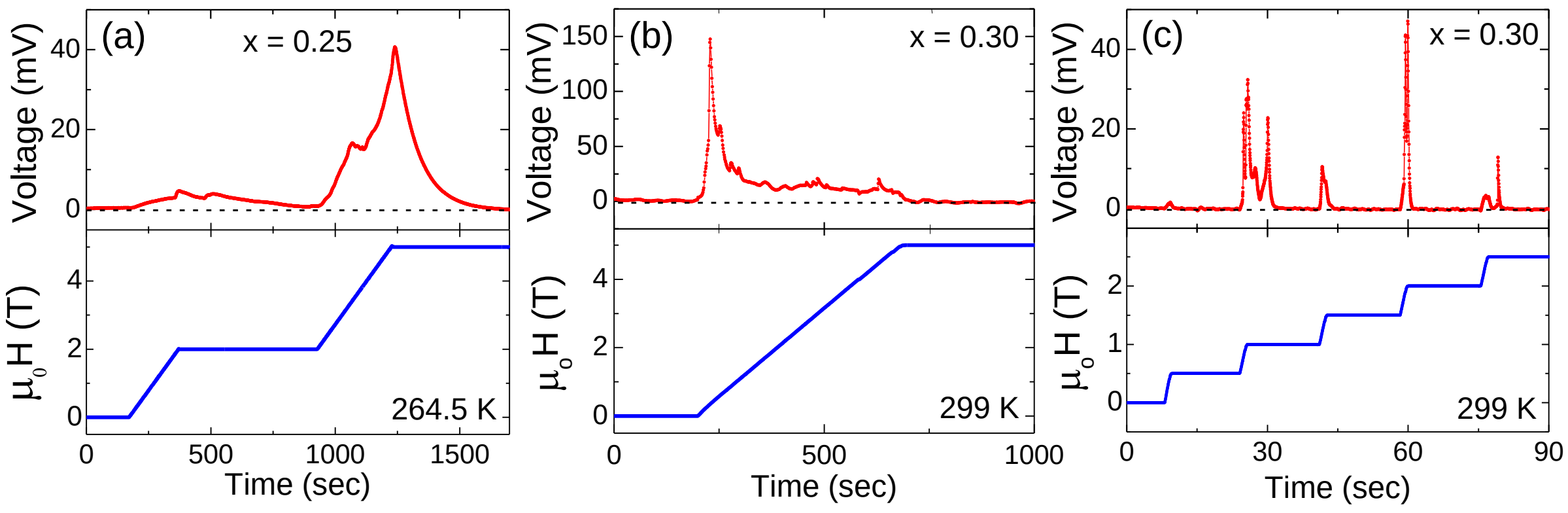

Figure 4: Output voltage measured in a Peltier cell associated with the heat released by the samples of $Ni_2Mn_{1-x}Cu_xGa_{0.84}Al_{0.16}$ (a) x = 0.25 and (b, c) x = 0.30 under magnetic field change at 264.5 K and 299 K, respectively. In (a), the magnetic field increases in two steps, from 0 to 2 T and 2 to 5 T, while in (b) the magnetic field increases continuously from 0 to 5 T. Another example of measurement, used to calculate ΔS as a function of magnetic field, is seen in panel (c), where various magnetic field steps were applied. This data allows us to obtain the total isothermal entropy change under a magnetic field variation.

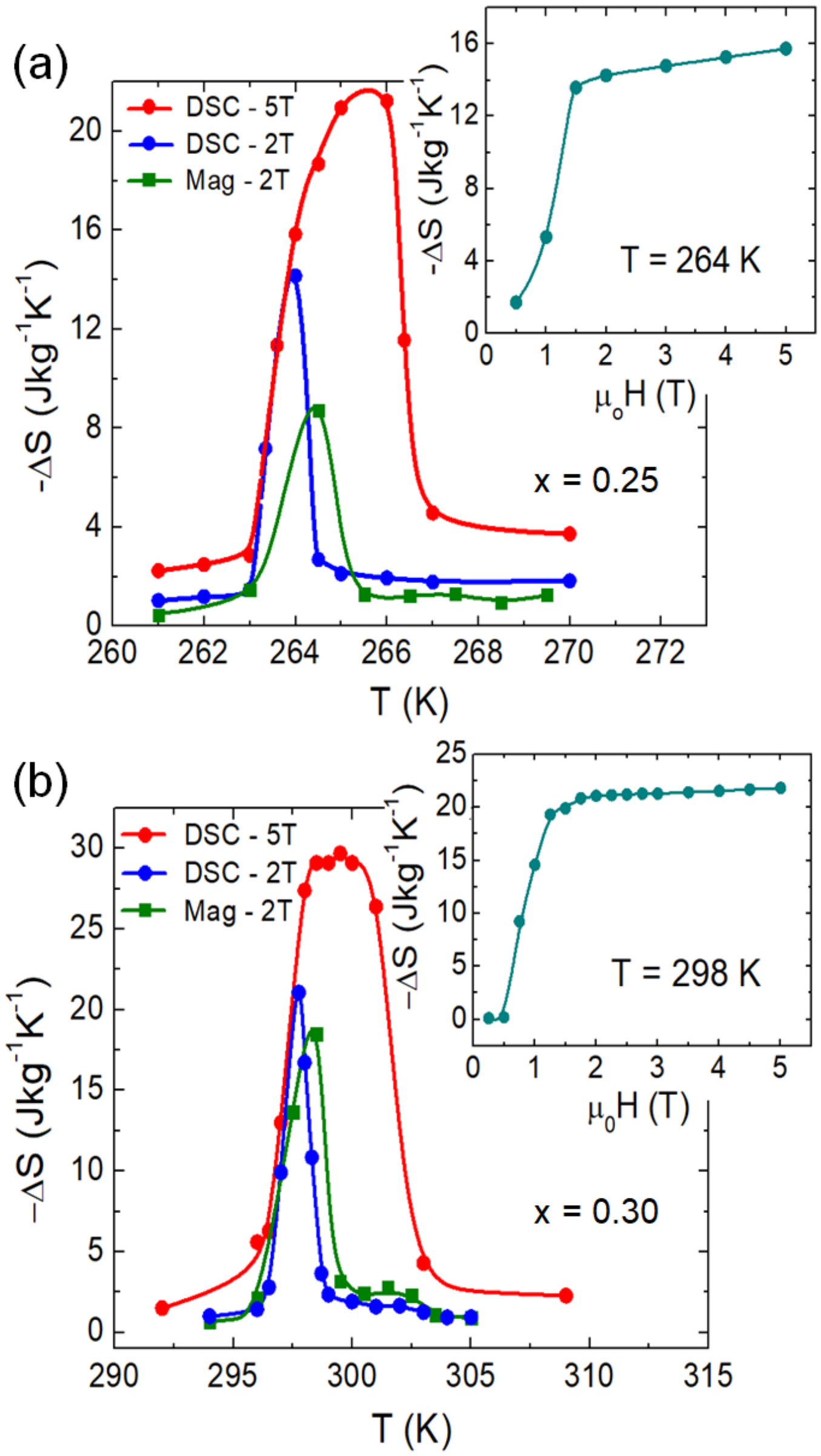

Figure 5: Temperature dependence of the total entropy change ΔS (measured with the Peltier calorimeter) and magnetic entropy change $\Delta S_M$ (obtained from magnetization data) of $Ni_2Mn_{1-x}Cu_xGa_{0.84}Al_{0.16}$, for (a) x = 0.25 and (b) x = 0.30, measured with 0 - 2 T and 0 - 5 T field change. The insets show the magnetic field dependence of the total entropy change ΔS.

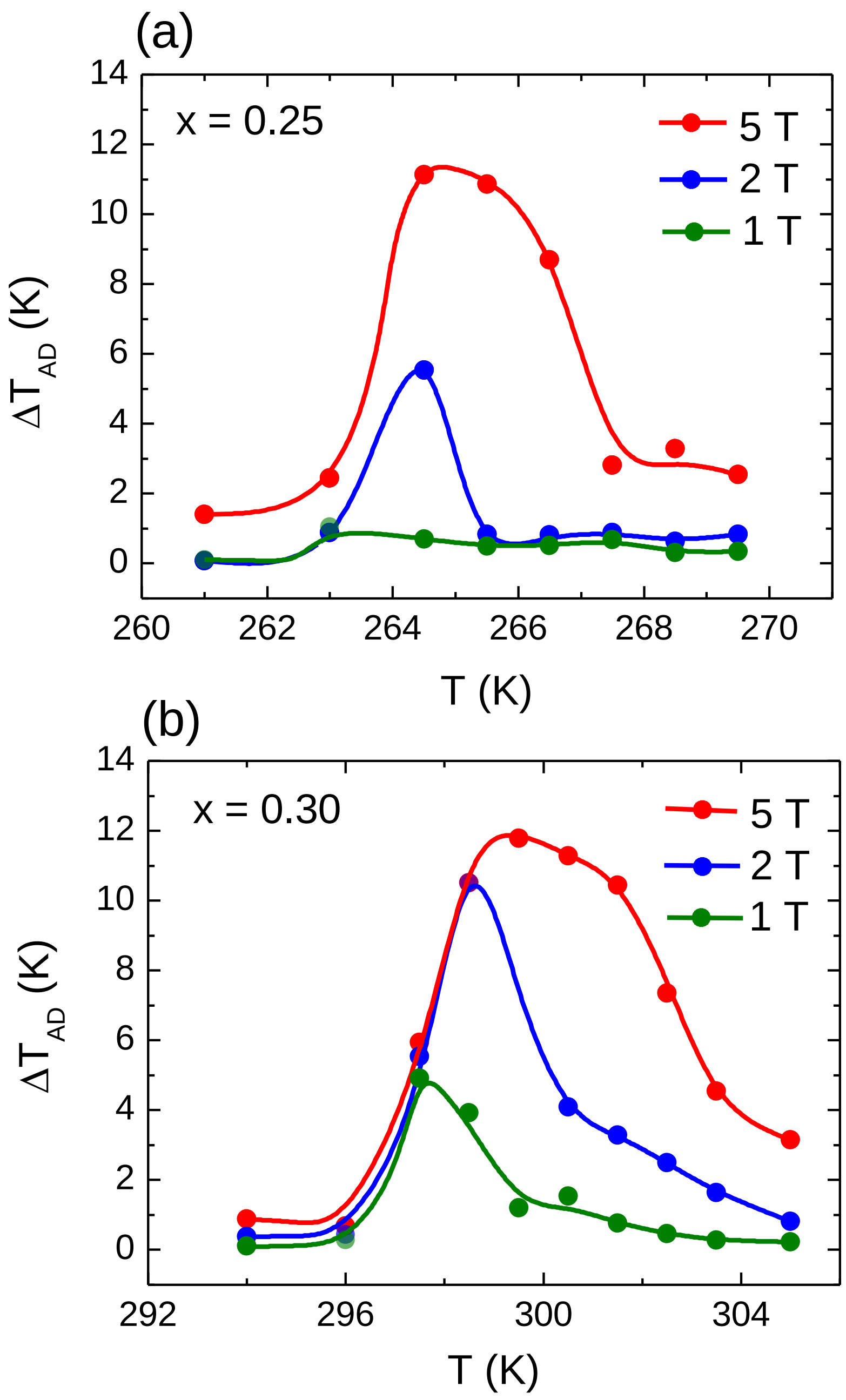

Figure 6: Adiabatic temperature change,$\Delta T_{AD}$ , as a function of temperature of $Ni_2Mn_{1-x}Cu_xGa_{0.84}Al_{0.16}$: (a) x = 0.25 and (b) x = 0.30 under 0 – 1 T, 0 - 2 T and 0 - 5 T magnetic field change. The values of $\Delta T_{AD}$ were calculated through specific heat and magnetization measurements.

Table 1: Total entropy change (ΔS) and refrigerant capacity (RC) of $Ni_2Mn_{1-x}Cu_xGa_{0.84}Al_{0.16}$ under 0 to 2 T and 0 to 5 T for x = 0.25 and x = 0.30 samples.

| Cu (x) | ΔS ($JKg^{-1}K^{-1}$) | | RC ($JKg^{-1}$) | |
|---|---|---|---|---|
| | 0 - 2 T | 0 - 5 T | 0 - 2 T | 0 - 5 T |
| 0.25 | -14 | -21 | 10 | 51 |
| 0.30 | -21 | -30 | 22 | 120 |

Table 2: Entropy change and adiabatic temperature change under 0 to 1 T, comparing our sample with selected compounds in the literature. Listed values for ΔS refer to either total or magnetic entropy change; details are given in each reference. Positive ΔS values are related to an inverse magnetocaloric effect.

| Composition | T (K) | ΔS ($Jkg^{-1}K^{-1}$) | $\Delta T_{AD}$ (K) |
|---|---|---|---|
| $Mn_1Fe_{0.95}P_{0.61}Si_{0.33}B_{0.06}$ [Ref. 29] | 267 | -19 | 1.3 |
| $Ni_2Mn_{0.7}Cu_{0.3}Ga_{0.84}Al_{0.16}$ [this work] | 298 | -15 | 4.8 |
| $Ni_{37.5}Co_{12.5}Mn_{35}Ti_{15}$ [Ref. 30] | 290 | 15 | -- |
| $LaFe_{11.47}Mn_{0.25}Si_{1.28}$-$H_{1.65}$ [Ref. 31] | 304 | -14 | 3 |
| $Mn_1Fe_{0.95}P_{0.605}Si_{0.33}B_{0.065}$ [Ref. 29] | 274 | -12.5 | 2.4 |
| $Mn_1Fe_{0.95}P_{0.6}Si_{0.33}B_{0.07}$ [Ref. 29] | 279 | -11.5 | 2.5 |
| $Gd_5Si_2Ge_2$ [Ref. 32] | 272 | -11 | -- |
| $Mn_{1.25}Fe_{0.70}P_{0.51}Si_{0.49}$ [Ref. 33] | 278 | -10.5 | 2.1 |
| $Mn_{0.85}Fe_{1.1}P_{0.62}Si_{0.32}B_{0.06}$ [Ref. 29] | 310 | -10.1 | -- |
| $Mn_{1.20}Fe_{0.80}P_{0.75}Ge_{0.25}$ [Ref. 33] | 282 | -10 | 1.7 |

**References**

[1] E. Liu, W. Wang, L. Feng, W. Zhu, G. Li, J. Chen, H. Zhang, G. Wu, C. Jiang, H. Xu & F. de Boer, Stable magnetostructural coupling with tunable magnetoresponsive effects in hexagonal phase-transition ferromagnets, Nat. Commun.**3**, 873 (2012).

[2] V.D. Buchelnikov and V.V. Sokolovskiy, Magnetocaloric effect in Ni-Mn-X (X = Ga, In, Sn, Sb) Heusler alloys, Phys. Metals Metallogr. **112**, 633 (2011).

[3] P.J. Webster, K.R.A. Ziebeck, S.L. Town, M.S. Peak, Magnetic order and phase transformation in $Ni_2MnGa$, Philos. Mag. B **49**, 295 (1984).

[4] R. Kainuma, K. Oikawa, W. Ito, Y. Sutou, T. Kanomatac and K. Ishidab, Metamagnetic shape memory effect in NiMn-based Heusler-type alloys, J. Mater. Chem.**18**, 1837 (2008).

[5] R. Kainuma, Y. Imano, W. Ito, Y. Sutou, H. Morito, S. Okamoto, O. Kitakami, K. Oikawa, A. Fujita, T. Kanomata and K. Ishida, Magnetic-field-induced shape recovery by reverse phase transformation, Nature **439**, 957 (2006).

[6] H. E. Karaca, I. Karaman, B. Basaran, Y. Ren, Y. I. Chumlyakov, and H. J. Maier, Magnetic field-induced phase transformation in NiMnCoIn magnetic shape-memory alloys - a new actuation mechanism with large work output, Adv. Funct. Mater. **19**, 983 (2009).

[7] K. Bhattacharya and R.D. James, The Material Is the Machine, Science **307**, 53 (2005).

[8] R.C. O'Handley, S.J. Murray, M. Marioni, H. Nembach, and S.M. Allen, Phenomenology of giant magnetic-field-induced strain in ferromagnetic shape-memory materials, J. Appl. Phys. **87**, 4712 (2000).

[9] M. Chmielus, X.X. Zhang, C.Witherspoon, D.C. Dunand and P. Müllner, Giant magnetic-field-induced strains in polycrystalline Ni–Mn–Ga foams, Nat. Mater.**8**, 863 (2009).

[10] A.A. Cherechukin, T. Takagi, M. Matsumoto, V.D. Buchelnikov, Magnetocaloric effect in $Ni_{2+x}Mn_{1-x}Ga$ Heusler alloys, Phys. Lett. A **326**, 146 (2004).

[11] S. Stadler, M. Khan, J. Mitchell, N. Ali, A. M. Gomes, I. Dubenko, A. Y.Takeuchi, and A. P. Guimarães, Magnetocaloric properties of $Ni_2Mn_{1-x}Cu_xGa$, Appl. Phys. Lett. **88**, 192511 (2006).

[12] Y. Liu, L. Luo, X. Zhang, H. Shen, J. Liu, J. Sun, N. Zu, Magnetostructural coupling induced magnetocaloric effects in Ni-Mn-Ga-Fe Microwires, Intermetallics **112**, 106538 (2019).

[13] S. K. Sarkar, Sarita, P.D. Babu, A. Biswas, V. Siruguri, M. Krishnan, Giant magnetocaloric effect from reverse martensitic transformation in Ni-Mn-Ga-Cu ferromagnetic shape memory alloys, J. Alloys Comp **670**, 281 (2016).

[14] M. Kataoka, K. Endo, N. Kudo, and T. Kanomata, H. Nishihara, T. Shishido and R. Y. Umetsu, M. Nagasako and R. Kainuma, Martensitic transition, ferromagnetic transition, and their interplay in the shape memory alloys $Ni_2Mn_{1-x}Cu_xGa$, Phys. Rev. B **82**, 214423 (2010).

[15] D. Zhao, T. Castán, A. Planes, Z. Li, W. Sun, and J. Liu, Enhanced caloric effect induced by magnetoelastic coupling in NiMnGaCu Heusler alloys: Experimental study and theoretical analysis, Phys. Rev. B **96**, 224105 (2017).

[16] H. Ishikawa, R. Y. Umetsu, K. Kobayashi, A. Fujita, R. Kainuma, and K. Ishida, Atomic ordering and magnetic properties in $Ni_2Mn(Ga_xAl_{1-x})$ Heusler alloys, Acta Mater. **56**, 4789 (2008).

[17] M. Acet, E. Duman, E. F. Wassermann, L. Mañosa, and A. Planes, Coexisting ferro- and antiferromagnetism in $Ni_2MnAl$ Heusler alloys, J. Appl. Phys. **92**, 3867 (2002).

[18] C. Salazar Mejía, A.M. Gomes, and L.A.S. de Oliveira, A less expensive NiMnGa based Heusler alloy for magnetic refrigeration, J. Appl. Phys. **111**, 07A923 (2012).

[19] A.A. Mendonça, L. Ghivelder, P.L. Bernardo, Hanlin Gu, R.D. James, L.F. Cohen, A,M. Gomes, Experimentally correlating thermal hysteresis and phase compatibility in multifunctional Heusler alloys, arXiv:2007.07485 [cond-mat.mtrl-sci].

[20] A.A. Mendonça, J.F. Jurado, S.J. Stuard, L.E.L. Silva, G.G. Eslava, L.F. Cohen, L. Ghivelder, and A.M. Gomes, Giant Magnetic Field Induced Strain in Ni2MnGa-based polycrystal, J. Alloys Comp. 738, 509 (2018).

[21] V. Basso, C. P. Sasso, and M. Küpferling, A Peltier cells differential calorimeter with kinetic correction for the measurement of Cp(H,T) and ΔS(H,T) of magnetocaloric materials, Rev. Sci. Instr. 81, 113904 (2010).

[22] V. V. Khovailo, V. Novosad, T. Takagi, D. A. Filippov, R. Z. Levitin, and A. N. Vasil'ev, Magnetic properties and magnetostructural phase transitions in Ni2+xMn1−xGa shape memory alloys, Phys. Rev. B **70**, 174413 (2004).

[23] P. J. Shamberger and F. S. Ohuchi, Hysteresis of the martensitic phase transition in magnetocaloric-effect Ni-Mn-Sn alloys, Phys. Rev. B **79**, 144407 (2009).

[24] G.J. Liu and J.R. Sun, J. Shen, B. Gao, H.W. Zhang, F.X. Hu, and B.G. Shen, Determination of the entropy changes in the compounds with a first-order magnetic transition, Appl. Phys. Lett. **90**, 032507 (2007).

[25] C. P. Bean, and D. S. Rodbell, Magnetic Disorder as a First-Order Phase Transformation, Phys. Rev. **126**, 104 (1962).

[26] V. Basso, Basics of the magnetocaloric effect, arXiv:1702.08347v1 [cond-mat.mtrl-sci] 27 (2017).

[27] M. Ghahremani, H. M. Seyoum, H. ElBidweihy, E. D. Torre, and L. H. Bennett, Adiabatic magnetocaloric temperature change in polycrystalline gadolinium – A new approach highlighting reversibility, Aip Advances **2**, 032149 (2012).

[28] M. Balli, D. Fruchart, D. Gignoux, and R. Zach, The colossal magnetocaloric effect in $Mn_{1-x}Fe_xAs$ : What are we really measuring?, Appl. Phys. Lett. **95**, 072509 (2009).

[29] F. Guillou, H. Yibole, G. Porcari, L. Zhang, N.H. van Dijk, and E. Brück. Magnetocaloric effect, cyclability and coefficient of refrigerant performance in the MnFe(P, Si, B) system, J. Appl. Phys **116**, 063903 (2014).

[30] H. Neves Bez, A. K. Pathak, A. Biswas, N. Zarkevich, V. Balema, Y. Mudryk, D. D. Johnson, V. K. Pecharsky. Giant enhancement of the magnetocaloric response in Ni-Co-Mn-Ti by rapid solidification, Acta Mater. **173**, 225 (2019).

[31] V. Basso, M. Küpferling, C. Curcio, C. Bennati, A. Barzca, M. Katter, M. Bratko, E. Lovell, J. Turcaud, and L. F. Cohen, Specific heat and entropy change at the first order phase transition of $La(Fe\text{-}Mn\text{-}Si)_{13}$-H compounds, J. Appl. Phys **118**, 053907 (2015).

[32] A. O. Pecharsky, K.A. Gschneidner Jr., and V.K. Pecharsky, The giant magnetocaloric effect of optimally prepared $Gd_5Si_2Ge_2$, J. Appl. Phys. **93**, 4722 (2003).

[33] H. Yibole, F. Guillou, L. Zhang, N.H. Van Dijk, and E Bruck, Direct measurement of the magnetocaloric effect in MnFe(P,X) (X = As, Ge, Si) materials, J. Phys. D: Appl. Phys. **47**, 075002 (2014).